\documentclass[twocolumn,prb,showpacs,preprintnumbers,amsmath,amssymb]{revtex4}

\usepackage{graphicx}

\begin{document}

\title{Broadband, noise-free optical quantum memory with neutral nitrogen-vacancy centers in diamond}

\author{E. Poem}
\email{e.poem1@physics.ox.ac.uk}
\author{C. Weinzetl}
\author{J. Klatzow}
\author{K. T. Kaczmarek}
\author{J. H. D. Munns}
\author{T. F. M. Champion}
\author{D. J. Saunders}
\author{J. Nunn}
\author{I. A. Walmsley}

\affiliation{
Clarendon Laboratory, University of Oxford, Parks Road, Oxford OX1 3PU, UK}

\date{\today}

\begin{abstract}
It is proposed that the ground-state manifold of the neutral nitrogen-vacancy center in diamond could be used as a quantum two-level system in a solid-state-based implementation of a broadband, noise-free quantum optical memory. The proposal is based on the same-spin $\Lambda$-type three-level system created between the two E orbital ground states and the A$_1$ orbital excited state of the center, and the cross-linear polarization selection rules obtained with the application of transverse electric field or uniaxial stress. Possible decay and decoherence mechanisms of this system are discussed, and it is shown that high-efficiency, noise-free storage of photons as short as a few tens of picoseconds for at least a few nanoseconds could be possible at low temperature.
\end{abstract}


\maketitle

\section{Introduction}\label{sect:1}

Quantum-optical memories (QOMs)~\cite{MemoryReviewSimon} are devices that can store optical quantum information, and retrieve it back on demand. QOMs are crucial for the synchronization of quantum optical devices and communication networks.~\cite{Synch_with_memories} This task requires QOMs that can sustain many storage attempts during storage time, and storage time long enough to allow for coordination and feed-forward.~\cite{Synch_with_memories}

The Raman QOM~\cite{Josh07,Raman_mem_th} is one type of QOM that meets both these requirements. It is based on the conversion of an input signal photon into a long lived material excitation, \emph{e.g.} a spin wave, with the use of a strong control pulse introduced into the medium with the signal photon. The energies of the signal and control differ by exactly the energy of the material excitation. The signal is retrieved by introducing another control pulse, converting the material excitation back into light. The storage time is the coherence time of the material excitation, and the minimal storage attempt time is the duration of the signal, \emph{i.e.} the inverse of its bandwidth, limited by the detuning of the signal and control from resonance, and by the energy of the material excitation.

This protocol has been recently implemented in Cs vapor, demonstrating the storage of $\sim$300~ps long photons for more than 1~$\mu$s.~\cite{Raman_mem1,Raman_mem2,Raman_mem3} One major drawback of the Cs vapor implementation is a relatively high noise level.~\cite{Raman_mem2,Raman_mem3} This results from the retrieval of spurious material excitations created by spontaneous Raman scattering of the control field due to the unavoidable coupling of the latter to the populated ground state.~\cite{no_circ_raman}

Here we propose the use of an ensemble of neutral nitrogen-vacancy centers (NV$^0$s) in diamond~\cite{Davies79} as an alternative, solid-state platform for a broadband Raman QOM. The NV$^0$ was previously proposed~\cite{Theor_NV0_4A2} as a platform for quantum information processing (QIP) based on a spin-coherent metastable excited state, inaccessible from the ground states by direct optical transition.~\cite{NV0_EPR}
Our proposal is based on a different set of NV$^0$ states, which, as shown below, demonstrate strong Raman coupling. This is the same-spin $\Lambda$-type three-level system formed by the two NV$^0$ orbital ground-states and the first optically-accessible excited state.

We analyze the level structure and show that the ground states could be manipulated by experimentally achievable external electric field or stress. This has two important implications. First, the ground-state splitting, and thus the acceptance bandwidth of the device could be controlled. Second, complete suppression of readout noise could be achieved by cross-linear polarization selection rules that emerge under the external field and enable the control field couple to only one of the ground states.

Note that in contrast to the Cs case, where the storage medium is a spin excitation, in our proposal it is the orbital degree of freedom of the ground states that is excited. Electronic orbitals are susceptible to lattice vibrations and local distortions. For orbitally degenerate ground states this is manifested through dynamic Jahn-Teller (DJT) distortions.~\cite{Davies79,Hepp14} Nevertheless, we show that at low temperatures, the coherence time should be at least a few nanoseconds, long enough for feed-forward.

The manuscript is organized as follows. In Sect.~\ref{sect:3} we analyze the level structure of NV$^0$ under electric field for the ideal case of no internal strain and a single NV$^0$ orientation, and calculate the Raman coupling and the noise suppression factor. The conditions required for a strong Raman coupling are discussed in Appendix~\ref{sect:2}. In Sect.~\ref{sect:5} we develop a scheme for implementing noise-free, Raman QOM using NV$^0$s in common diamond samples containing NV$^0$s of all orientations, and estimate the expected memory efficiency. In Sect.~\ref{sect:4} we discuss the influence of random strain and phonon coupling on the ground-state manifold, and estimate the expected noise suppression factor and memory time for realistic experimental conditions. Spin fluctuations, charge fluctuations, and intersystem crossing are discussed as well. Low-temperature spectral measurements of the NV$^0$ fluorescence, relevant to the discussion, are presented in Appendix~\ref{app:1}. Finally, in Sect.~\ref{sect:6} we conclude.

\section{NV$^0$ -- an orbital $\Lambda$-system}\label{sect:3}
\begin{figure}[tbh]
\includegraphics[width=0.46\textwidth]{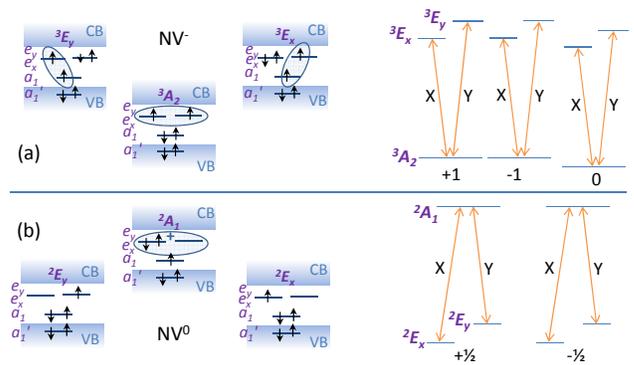}
  \caption{
(a) [(b)] Electronic configurations (left) and schematic level structure (right) of the NV$^-$ (NV$^0$) under electric field along the $\hat{x}$ direction in the NV coordinate system (see text). VB (CB) denotes the valence (conduction) band. An upward- (downward-) pointing arrow represents an electron with spin $\tfrac{1}{2}$ (-$\tfrac{1}{2}$). The ellipses in (a) represent spin-triplet configurations. Only one of the three states is presented. The ellipse in (b) represents a symmetric superposition of both electrons occupying the E$_x$ state or the E$_y$ state. Only the spin $\tfrac{1}{2}$ states are presented. The double-headed arrows on the right represent allowed optical transitions. The transition dipole direction is specified next to each arrow.}
  \label{fig1}
\end{figure}

The nitrogen-vacancy (NV) center in diamond has two charge states: neutral, NV$^0$, and negatively charged, NV$^-$. Fig.~\ref{fig1} presents the ground and optically excited states of the NV$^-$ [panel (a)] and the NV$^0$ [panel (b)], and the allowed optical transitions between them, under an external electric field or stress large enough such that spin mixing is negligible (see below).

The NV$^-$ ground state manifold is composed of three different spin states of the same electronic orbital. This is one of the reasons for which this system is coherent even at room temperature,~\cite{Wrachtrup09} which enables its application for QIP,~\cite{Hanson06,Hanson13} metrology,~\cite{Taylor08,Wrachtrup09} and more. It was also recently proposed as a platform for Raman QOM.~\cite{Simon14}

However, from Fig.~\ref{fig1}(a) it is clear that without spin mixing, the NV$^-$ states form three independent V-systems, each of a different spin. This means that Raman coupling between different ground states of the NV$^-$ depends on spin mixing interactions (SMIs). As we show in Appendix~\ref{sect:2}, for detunings larger than the SMIs, the Raman coupling goes down \emph{quadratically}. Therefore, the maximal detuning, and thus the acceptance bandwidth of an NV$^-$ Raman QOM, would be limited by the strength of the SMIs.

SMIs can be transverse spin-orbit (SO), transverse external magnetic field, or transverse spin-spin interactions. In the NV$^-$, due to the low symmetry, the SO does not have a transverse part within the excited states manifold.~\cite{Hollenberg11,Maze11} Transverse spin-spin interactions can mix excited states of different spins when they are brought together by external magnetic,~\cite{Awschalom13,Awschalom14} electric,~\cite{Tamarat08,Acosta12} or stress~\cite{Santori06,Acosta13,Batalov09} fields. External magnetic field with a small transverse component can do the same for the ground states.~\cite{gs_mw_lambda} In all those cases, however, the SMI is on the order of 1~GHz or less.~\cite{Awschalom14}

For the $\sim$30~GHz detuning required to avoid the inhomogeneous broadening of the NV$^-$ transition energy,~\cite{Santori06,Acosta13} and accommodate signal pulses of 1~ns or less, the Raman coupling would be about two orders of magnitude smaller than in an equivalent system where the SMI is much larger than the detuning, or where SMIs are not required for Raman coupling.

As shown in Fig.~\ref{fig1}(b), the states of the NV$^0$ form two independent orbital $\Lambda$-systems, one for each spin component. Here the Raman coupling is not limited by SMIs. We discuss this structure in more detail below.

\subsection{Level structure and optical transitions under electric field, neglecting random strain}
For the ideal case where internal strain is negligible, the NV$^0$ Hamiltonian under an external electric field (or equivalently, external uniaxial stress~\cite{Davies79}) reads,~\cite{Hollenberg11,Maze11}
\begin{equation}\label{eq5}
\begin{array}{lcl}
\hat{H}&=&\hat{H}_0+\lambda_{\parallel}\hat{L}_z\hat{S}_z+\lambda_{\perp}\left(\hat{L}_x\hat{S}_x+\hat{L}_y\hat{S}_y\right)\\
& &+\hat{H}_{DJT}-\hat{\vec{d}}\cdot\vec{F},
\end{array}
\end{equation}
where $\hat{H}_0$ includes the single-electron energies and the Coulomb interaction, $\lambda_{\parallel(\perp)}$ is the longitudinal (transverse) SO coupling energy, $\hat{H}_{DJT}$ is the DJT Hamiltonian,~\cite{Hepp14} $\hat{\vec{d}}$ is the electric dipole vector operator,~\cite{Hollenberg11,Maze11} and $\vec{F}$ is the electric field vector.
The axes are chosen such that $\hat{z}$ points from the nitrogen atom to the vacancy and $\hat{x}$ lies in one of the three vertical reflection planes of the NV center.
No spin-spin interaction is included since, as shown in Fig.~\ref{fig1}(b), there is only one open-shell electron in all the relevant states.

Due to their large separation, we treat the ground and excited states separately. The excited states can be spanned by the basis $\{\mbox{A}_{1,\uparrow},\mbox{A}_{1,\downarrow}\}$. They include one non-degenerate orbital, and are not affected by DJT distortions. They also have no angular-momentum, and only longitudinal dipole moment. Therefore, the Hamiltonian within this manifold is proportional to a unit matrix,
\begin{equation}\label{eq6}
\hat{H}_{es}=(\varepsilon_{es}-d_{\parallel}F_z)I_2,
\end{equation}
where $\varepsilon_{es}$=521.4~THz is the zero-field excited-state energy with respect to the mean ground-state energy, $F_z$ is the $\hat{z}$ component of the electric field, and $I_2$ is a 2$\times$2 unit matrix. Since the field dependence of the excited-state manifold is relatively weak,~\cite{Davies79} and does not lift any degeneracy, it will be neglected in the following discussion.

The ground state manifold can be spanned by the basis $\{\mbox{E}_{x,\uparrow},\mbox{E}_{x,\downarrow},\mbox{E}_{y,\uparrow},\mbox{E}_{y,\downarrow}\}$. Within this manifold, the transverse SO interaction vanishes,~\cite{Tamarat08,Batalov09} but the DJT energies may not be zero. The Hamiltonian then becomes,~\cite{Hollenberg11,Maze11}
\begin{equation}\label{eq7}
\begin{array}{lcl}
H_{gs}&=&\frac{\lambda_{\parallel}}{2}\left(
       \begin{array}{cc}
           0_2 & -i\sigma_z\\
           i\sigma_z & 0_2 \\
         \end{array}
       \right)
       +\left(\Upsilon_x-d_{\perp}F_x\right)\left(
         \begin{array}{cc}
           I_2 & 0_2\\
           0_2 & -I_2\\
         \end{array}
       \right)\\& &+\left(\Upsilon_y-d_{\perp}F_y\right)\left(
         \begin{array}{cc}
           0_2 & I_2\\
           I_2 & 0_2\\
         \end{array}
       \right),
\end{array}
\end{equation}
where $\sigma_z$ is the third Pauli matrix, $0_2$ is a 2$\times$2 zero matrix, $F_{x(y)}$ is the $\hat{x}$ ($\hat{y}$) component of the electric field, $\Upsilon_x$ and $\Upsilon_y$ are the DJT energies, and the mean energy of the E manifold at zero field was set to zero.

The longitudinal SO interaction energy and the transverse electric dipole moments are estimated to be $\lambda_{\parallel}\approx4.3$~GHz and $d_{\perp}\approx5$~GHz$\cdot\mu\mbox{m}/\mbox{V}$, respectively. Note that since, to the best of our knowledge, there are no direct measurements of these coefficients, they are estimated using the values measured for the excited E manifold of the NV$^-$,~\cite{Batalov09,Tamarat06} corrected by the ratio between the measured strain energy coefficients of the NV$^0$ and the NV$^-$.~\cite{DaviesHamer76,Davies79} This correction is applied in order to account for the different influence of DJT distortions on the two charge states.

The values of the DJT energies are also unknown. For the excited E states of the NV$^-$, the measured level structure could be fully explained assuming zero DJT energies.~\cite{Batalov09,Acosta12,Awschalom14} While this does not necessarily mean that these energies are zero also for the NV$^0$ ground states, an upper bound could be set by noting that together with the SO energy, the DJT energies would split the NV$^0$ zero-phonon line (ZPL) at zero external fields by $S_0=\sqrt{4\Upsilon^2+\lambda_{\parallel}^2}$, where $\Upsilon=\sqrt{\Upsilon_x^2+\Upsilon_y^2}$. We have performed low temperature measurements on an NV$^0$ ensemble (see Appendix~\ref{app:1}), and could bound any possible splitting to below 24~GHz. Given the estimated value of $\lambda_{\parallel}$, we set an upper limit of 12~GHz on $\Upsilon$.
\begin{figure}[tbh]
\includegraphics[width=0.5\textwidth]{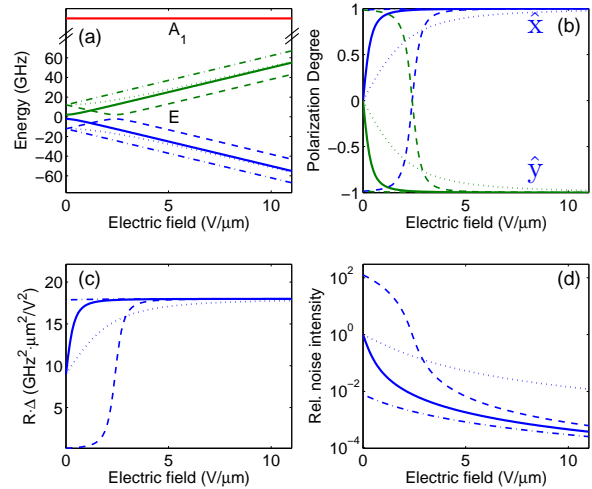}
  \caption{(a) NV$^0$ energy levels vs. transverse electric field. The NV axis is along $\left[111\right]$ ($\hat{z}$). The field is along $\left[\bar{1}\bar{1}2\right]$ ($\hat{x}$). The solid line presents the case of $\Upsilon$=0. The dashed, dotted, and dashed-dotted lines present the case of $\Upsilon$=12~GHz, for $\alpha$=$0^{\circ}$, $\pm90^{\circ}$, and $180^{\circ}$, respectively. The same convention is used also in (b)-(d). (b) Linear polarization degree [equals 1 (-1) for $\hat{x}$ ($\hat{y}$) polarization] of the optical transitions related to the lower (blue/ dark gray lines) and the higher (green/light gray lines) ground state. (c) The product of the Raman coupling and the detuning, detuning-independent in the large-detuning regime, in cross-linear polarizations [higher (lower) energy light polarized along $\hat{x}$ \mbox{($\hat{y}$)}]. (d) Relative probability of $\hat{x}$-polarized light to couple to the higher ground state.}
  \label{fig2}
\end{figure}

The energies of the ground and excited states, as functions of an external electric field applied along the $\hat{x}$ direction ($\hat{x}\parallel\left[\bar{1}\bar{1}2\right]$~\cite{comment1} for an NV aligned along $\hat{z}\parallel\left[111\right]$), are presented in Fig.~\ref{fig2}(a).
The solid line presents the case of $\Upsilon$=0. The dashed, dotted, and dashed-dotted lines present the case of $\Upsilon$=12~GHz, for $\alpha$=0$^{\circ}$, $\pm90^{\circ}$, and 180$^{\circ}$, respectively. Here $\alpha=\tan^{-1}(\Upsilon_y/\Upsilon_x)$.
At zero external field, the SO and DJT interactions mix the E$_x$ and E$_y$ orbitals. In the case of no DJT interaction, the states would be the total angular momentum projection states, \mbox{$\mbox{E}_{\pm\frac{1}{2}}=(\mbox{E}_{x,\uparrow(\downarrow)}\mp i\mbox{E}_{y,\uparrow(\downarrow)})/\sqrt{2}$}, and \mbox{$\mbox{E}_{\pm\frac{3}{2}}=(\mbox{E}_{x,\uparrow(\downarrow)}\pm i\mbox{E}_{y,\uparrow(\downarrow)})/\sqrt{2}$}. With the DJT interaction, other E$_x$ and E$_y$ superpositions will form, depending on $\Upsilon_x$, $\Upsilon_y$, and $\lambda_{\parallel}$. For external fields larger than $S_0/d_g\lesssim5$~V/$\mu$m, the states closely approach $\{\mbox{E}_{x,\uparrow},\mbox{E}_{x,\downarrow},\mbox{E}_{y,\uparrow},\mbox{E}_{y,\downarrow}\}$.

As shown in Fig.~\ref{fig1}(b), the A$_1$ excited-state orbital is optically connected to $both$ E$_x$ and E$_y$ ground-state orbitals, creating a $\Lambda$-system. This enables non-vanishing first-order Raman coupling (FORC) between the two orbital ground states (see Appendix~\ref{sect:2}.)

Furthermore, in this basis, ground states are coupled to the excited state in \emph{orthogonal linear polarizations}. The dipole transition matrices between the excited states, \mbox{$\{\mbox{A}_{1,\uparrow},\mbox{A}_{1,\downarrow}\}$}, and the ground states, $\{\mbox{E}_{x,\uparrow},\mbox{E}_{x,\downarrow},\mbox{E}_{y,\uparrow},\mbox{E}_{y,\downarrow}\}$, for the three dipole components, are,
\begin{equation}\label{eq8}
D^x=\frac{d_{ge}}{\sqrt{2}}\left(
       \begin{array}{c}
           I_2 \\
           0_2
         \end{array}
       \right), D^y=\frac{d_{ge}}{\sqrt{2}}\left(
       \begin{array}{c}
           0_2 \\
           I_2
         \end{array}
       \right), D^z=\left(
       \begin{array}{c}
           0_2 \\
           0_2
         \end{array}
       \right).
\end{equation}
The value of $d_{ge}$, estimated from the lifetime of the NV$^0$ (\mbox{$\sim$20 ns~\cite{NV0_lifetime}}) using the formula derived in Ref.~\onlinecite{Simon14}, is about 6~GHz$\cdot\mu\mbox{m}/\mbox{V}$.

Fig.~\ref{fig2}(b) shows the linear polarization degrees of the optical transitions in the NV$^0$ coordinate system, \mbox{$(P_x-P_y)/(P_x+P_y)$}, calculated according to Eq.~(\ref{eq7}) and Eq.~(\ref{eq8}), as functions of the electric field, applied along the $\hat{x}$ direction. Here $P_{x(y)}$ is the absorbtion probability of $\hat{x}$ ($\hat{y}$) polarized light.

Fig.~\ref{fig2}(c) shows the resulting Raman coupling - detuning product, R$^{xy}_{E_1E_2}\bar{\Delta}$, detuning-independent in the large-detuning regime [see \mbox{Eq.~(\ref{eq1})} in \mbox{Appendix~\ref{sect:2}}], for two ground states of different orbitals and the same spin, in orthogonal linear polarizations ($\hat{x}$-polarized control and $\hat{y}$-polarized signal).

For quantum memory operation with a single photon as the signal, polarization alone is usually not sufficient for separating the strong control from the weak signal. Energy selective filtering is then used, and the energetic separation between the control and signal fields, tuned to the energy of the material excitation (the separation between the two ground-state levels), sets the maximum bandwidth of the memory.~\cite{Raman_mem1,Raman_mem2}
As shown in Fig.~\ref{fig2}(a), energy splittings in excess of 50~GHz are achievable with experimentally demonstrated electric fields ($\sim$5~V/$\mu$m)~\cite{Tamarat06,Tamarat08,Acosta12} and/or moderate compressive uniaxial stress ($\sim$50~MPa).~\cite{Davies79} Thus, the bandwidth of the proposed QOM could be as large as 20~GHz, permitting storage of 15~ps pulses. Such large bandwidths make the NV$^0$ system suitable for interfacing directly with parametric down-conversion photon sources.~\cite{Raman_mem3,Diamond_phonon_memory2}

The cross-linearly polarized selection rules induced by the external field
may additionally provide a way to suppress the readout noise encountered in alkali-metal vapor Raman QOMs,~\cite{Raman_mem2} as a polarized control field would couple mostly to one of the ground states, considerably suppressing the probability of creating and retrieving a material excitation in the absence of the signal field. This is shown in Fig.~\ref{fig2}(d), which presents the probability of an $\hat{x}$-polarized optical field to couple to the higher ground state, as a function of the applied external electric field. For electric field- or stress-induced splitting, $S$, much larger than the zero-field splitting, $S_0$, this probability, which is also the noise suppression factor, is approximately given by,
\begin{equation}\label{eq9}
P\approx\frac{\lambda^2_{\parallel}+4\Upsilon_y^2}{4(S-2\Upsilon_x)^2}.
\end{equation}
Given a fixed value of $\Upsilon$, the maximum value of $P$ is obtained, under the same approximation, for $\alpha$=$\pm90^{\circ}$. Therefore, for \mbox{$S=50$~GHz}, $P$ is at most 1/16, significantly reducing the readout noise.

Finally, a cross-linearly polarized $\Lambda$-system enables technically simple optical preparation and detection of the orbital quantum state of the electrons. Optical preparation can be achieved by resonant polarized excitation, coupled to just one of the ground states, and the consequent pumping of population to the uncoupled state. Optical detection can be achieved by resonant polarized excitation, projecting the population of the coupled state onto the excited state, followed by the measurement of the resulting excited state population by, \emph{e.g.}, measuring the resulting fluorescence (either at the ZPL, or at the phonon side band).

\section{Quantum-optical memory using NV$^0$s in a (001) diamond sample}\label{sect:5}
For an ensemble-based quantum memory, ideally one would like to have an ensemble of [111]-oriented NVs. Though the growth of diamond samples with a single NV orientation is under development, and significant advancements towards this goal have been very recently made,~\cite{directional_growth_germans,directional_growth_french} such samples are not readily available yet. Nevertheless, as will now be shown, it is possible to obtain a practically noise-free ensemble quantum memory also with the use of common, (001)-faceted samples containing all NV orientations.

\subsection{Proposed experimental configuration}
\begin{figure}[tbh]
\includegraphics[width=0.5\textwidth]{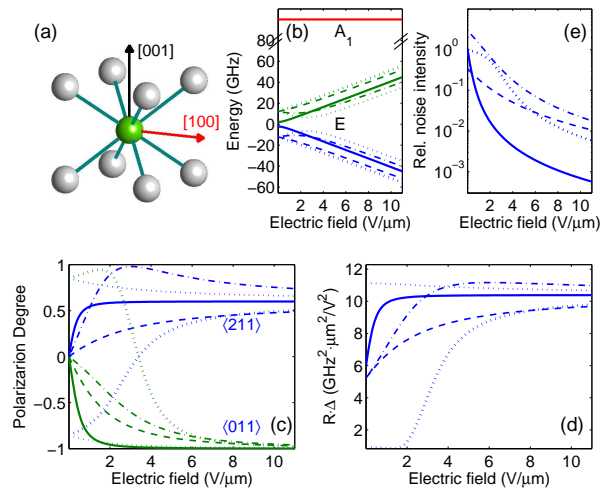}
\caption{(a) All possible orientations of the NV center. The nitrogen atom is in the middle.
The direction of the applied field (optical axis) is denoted by a red, horizontal (black, vertical) arrow. (b) NV$^0$ energy levels vs. electric field along [100]. The solid line presents the case of $\Upsilon$=0. The dashed, dotted, thick dotted, and dashed-dotted lines present the case of $\Upsilon$=12~GHz, for $\alpha$=$0^{\circ}$, $90^{\circ}$, $-90^{\circ}$, and $180^{\circ}$, respectively.
(c) The allowed transitions' $\left[100\right]$-$\left[010\right]$ linear polarization degree. The blue/dark gray (green/light gray) lines present the polarization of the transition coupled to the lower (higher) ground state.
(d) Raman coupling [the higher (lower) energy light is polarized along $\left[100\right]$ ($\left[010\right]$)].
(e) Relative probability of the higher-energy light to couple to the higher energy ground state, summed over all orientations.
}
\label{fig3}
\end{figure}

In order to utilize NVs of all orientations in a (001) sample, we propose to apply the electric field along the $\left[100\right]$ direction, and align the optical axis along the $\left[001\right]$ direction. Fig.~\ref{fig3}(a) presents all the possible orientations of the NV center with respect to these directions. Due to symmetry, the magnitude of the electric field projections on the planes perpendicular to the NV $\hat{z}$ axes are all the same. Fig.~\ref{fig3}(b) presents the energies of the NV$^0$ states as functions of the applied field. The solid line presents the case of $\Upsilon$=0, while the dashed, dotted, thick dotted, and dashed-dotted lines present the case of $\Upsilon$=12~GHz for $\alpha$=$0^{\circ}$, $90^{\circ}$, $-90^{\circ}$, and
$180^{\circ}$, respectively. Due to the equal projection magnitudes, for $\Upsilon$=0, NVs of all orientations show the same energy-field dependence. This is also true when $\Upsilon_x\neq0$ but only for $\Upsilon_y$=0. For $\Upsilon_y\neq0$ (dotted lines), in half of the orientations the cases of $\Upsilon_y>0$ and $\Upsilon_y<0$ switch (not shown).

The high-field transition-dipole projection on the (001) plane is also the same for all orientations. As shown in Fig.~\ref{fig3}(c), for a high-enough field, the dipole moment of the high-energy transition (solid line), asymptotically aligns along the projection direction of the applied field on the $x-y$ plane, [$\pm2\pm1\pm1$], and for all orientations, its linear polarization degree relative to the [100] and [010] axes goes to 0.6. The lower-energy transition dipole moment asymptotically aligns along [0$\pm1\pm1$], a direction in the $x-y$ plane that is perpendicular to the projection of the applied field. It therefore contains almost no component along the direction of the applied field, and its linear polarization degree tends to -1. The Raman coupling for a higher (lower) frequency control (signal) field polarized along $\left[100\right]$ ($\left[010\right]$) is presented in Fig.~\ref{fig3}(d). Here too, for half of the orientations, the cases of $\Upsilon_y>0$ and $\Upsilon_y<0$ switch (not shown). Nevertheless, for all orientations, the Raman coupling eventually saturates at 1/$\sqrt{3}$ of the asymptotical value calculated for the case where both the NV and the optical axis are oriented along the [111] direction [Fig.~\ref{fig2}(c)].

Furthermore, as shown in Fig.~\ref{fig3}(e), the total probability for readout noise, including all NV orientations, is still considerably suppressed. For a splitting of 50~GHz, the noise suppression factor is at most 1/20.

\begin{figure}[tbh]
\includegraphics[width=0.46\textwidth]{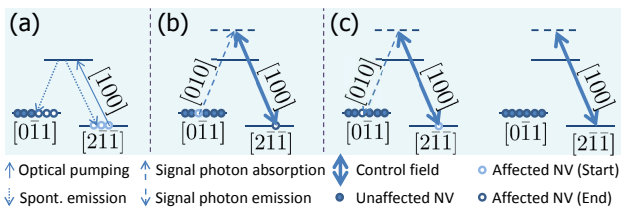}
  \caption{
(a) Optical pumping, (b) memory read-in, and (c) memory readout, for an NV oriented along $\left[111\right]$ under electric field along [100].
The transition dipole direction is stated next to each ground state. The polarization directions of the optical fields are indicated next to the relevant arrows. The right diagram in (c) corresponds to a readout attempt where no photon was initially stored. The control pulse couples only to one ground state, and no noise photon is emitted. The same external field directions induce the same effects for all other NV orientations as well (not shown).}
\label{fig4}
\end{figure}

Finally, optical pumping and detection would still be effective: for a $\left[100\right]$ polarized pump, coupled only to the lower ground state, all the NVs will be pumped to the higher ground state, and upon [100] polarized excitation, the resulting fluorescence or absorption would be proportional to the population in the lower ground state.
The optical pumping, memory read-in, and memory readout, are schematically illustrated in Fig.~\ref{fig4}.

\subsection{Expected memory efficiency}
The efficiency of the memory can be estimated from the relative Raman coupling strength, $\mathcal{R}$, defined as,~\cite{Josh07,Raman_mem_th}
\begin{equation}\label{eq21}
\mathcal{R}\approx\sqrt{\frac{\pi^2h}{\varepsilon_0^2c}}\cdot\sqrt{nE_C}\cdot\frac{R\cdot\Delta}{\lambda_C\Delta},
\end{equation}
where $R\cdot\Delta$ is the calculated detuning-independent product of the Raman coupling coefficient and the detuning [in units of Hz$^2$m$^2$/V$^2$, see Fig.~\ref{fig3}(d)], $E_C$ is the control pulse energy, $\lambda_C$ ($\Delta$) is its wavelength (detuning), $n$ is the NV$^0$ density, $\varepsilon_0$ is the permittivity of vacuum, $h$ is the Plank constant, and $c$ is the speed of light in vacuum. It is assumed that the Rayleigh range of the control and signal beams is matched to the sample length. For control pulses of 10~nJ, NV$^0$ density of 10$^{16}$~cm$^{-3}$, and a detuning of 100~GHz, significantly larger than the measured low-temperature inhomogeneous broadening (see Appendix~\ref{app:1}), $\mathcal{R}\sim$1 is calculated, which implies a $\sim$25\% total memory efficiency.~\cite{Josh07,Raman_mem_th} The efficiency may be significantly improved by the use of waveguides to increase the distance along which a high control intensity can be maintained.~\cite{Fibre_mem} In this case, in Eq.~(\ref{eq21}) the wavelength, $\lambda_C$, should be replaced by the geometric ratio $d^2/L$, where $d$ is the width of the waveguide, and $L$ is its length. For a \mbox{1~$\mu$m~$\times$~1~$\mu$m~$\times$~1~mm} waveguide, $\mathcal{R}>50$ is calculated even for a control pulse energy as low as 0.1~nJ, indicating a total memory efficiency that exceeds 99\%.~\cite{Josh07,Raman_mem_th}\\

\section{Decoherence mechanisms}\label{sect:4}
The use of a same-spin, orbital two-level system in the solid state as a quantum bit comes with the cost of increased sensitivity of the system to lattice distortions and vibrations compared to a same-orbital, different-spin system. The orbital states might also be influenced by spin fluctuations, through the SO coupling. Furthermore, the system might be driven into optically inactive, ``dark'' states. In this section we analyze the influence of these effects on the life and coherence times of the NV$^0$ ground-state manifold.

\subsection{Random local strain}\label{sect:4a}
\subsubsection{Inhomogeneous broadening}
Random local strain will inhomogeneously broaden the energy levels in the NV$^0$ system. The broadening of the optical energy difference is not a major limiting factor since the FORC in this system does not vanish, and one could detune away from the broadened transition energy with little reduction in Raman coupling. The ground-state splitting will also be broadened, leading to increased dephasing of ground-state orbital coherence. In principle, if the broadening is smaller than the splitting, this decoherence can be counteracted by spin-echo techniques.~\cite{SpinEchoReview}

In order to estimate the relative magnitudes of the strain broadening on the different energy levels and their dependence on the applied external electric field, we write down the linear strain Hamiltonian, which has a similar form to that of the electric field Hamiltonian. For the excited states it reads,~\cite{Davies79}
\begin{equation}\label{eq10}
H^{(e)}_{es}=(\epsilon_{A_1}e_{A_1}+\epsilon'_{A_1}e'_{A_1})I_2,
\end{equation}
where $e_{A_1}=e_{zz}$ and $e'_{A_1}=e_{xx}+e_{yy}$ are the two possible A$_1$ deformation modes that are linear in the strain tensor components, $e_{ij}$ (where $i,j\in\{x,y,z\}$), and \mbox{$\epsilon_{A_1}\approx 192$~THz} and \mbox{$\epsilon'_{A_1}\approx -483$~THz} are the corresponding strain energies.~\cite{Davies79}

For the ground states, the strain Hamiltonian reads,~\cite{Davies79}
\begin{equation}\label{eq11}
\begin{array}{lcl}
H^{(e)}_{gs}&=&(\epsilon_{E}e_{E_x}+\epsilon'_{E}e'_{E_x})\left(
         \begin{array}{cc}
           I_2 & 0_2\\
           0_2 & -I_2\\
         \end{array}
       \right)\\
       & &+(\epsilon_{E}e_{E_y}+\epsilon'_{E}e'_{E_y})\left(
         \begin{array}{cc}
           0_2 & I_2\\
           I_2 & 0_2\\
         \end{array}
       \right),
\end{array}
\end{equation}
where \mbox{$\{e_{E_x},e_{E_y}\}=\{e_{xx}-e_{yy},2e_{xy}\}$} and \mbox{$\{e'_{E_x},e'_{E_y}\}=\{2e_{xz},2e_{yz}\}$} are the two possible E deformation-mode pairs that are linear in the strain, and \mbox{$\epsilon_{E}\approx -600$~THz} and \mbox{$\epsilon'_{E}\approx 360$~THz} are the corresponding strain energies.~\cite{Davies79}

Diagonalizing the total Hamiltonian, including strain, the eigen-energies of the ground-states with respect to the excited states take the form,
\begin{equation}\label{eq12}
\varepsilon_{1,2}=A\pm\sqrt{B^2+C^2+D^2},
\end{equation}
where,
\begin{equation}\label{eq13}
 \begin{array}{lcl}
 A&=&-\varepsilon_{es}+d_{\parallel}F_z-\epsilon_{A_1}e_{A_1}-\epsilon'_{A_1}e'_{A_1},\\
 B&=&\Upsilon_x-d_{\perp}F_x+\epsilon_{E}e_{E_x}+\epsilon'_{E}e'_{E_x},\\
 C&=&\Upsilon_y-d_{\perp}F_y+\epsilon_{E}e_{E_y}+\epsilon'_{E}e'_{E_y},\\
 D&=&\frac{\lambda_{\parallel}}{2}.
 \end{array}
\end{equation}

In the following, we analyze the simple case of an external field applied along the $\hat{x}$-direction, as in Sect.~\ref{sect:3}. The same analysis can also be applied for the case presented in Sect.~\ref{sect:5}, yielding similar conclusions.

For a large field applied along the $\hat{x}$-direction, inducing a ground-state splitting much larger than any zero-field splitting, the energies are approximately $\varepsilon_{1,2}\approx A\pm B$, and in the presence of random strain, the distribution of ground-state energy splittings would be related mostly to the distributions of E$_x$ type strains,
\begin{equation}\label{eq14}
\begin{array}{l}
\delta S\equiv\delta(\varepsilon_1-\varepsilon_2)\approx2\delta B\\
\approx2\sqrt{\epsilon^2_{E}\delta e^2_{E_x}+\epsilon'^2_{E}\delta e'^2_{E_x}}\approx2\delta e\sqrt{\epsilon^2_{E}+\epsilon'^2_{E}},
\end{array}
\end{equation}
where in the last step we assumed an isotropic random strain distribution with a standard deviation of $\delta e$.

One simple way to estimate the random strain in a given sample is by measuring the broadening of the NV$^0$ zero-phonon line in the absence of external fields and at low temperature (where the broadening due to dynamical distortions is quenched -- see Sect.~\ref{sect:4b} below). In this case, for $\Upsilon\neq0$, the width of the distribution of transition energies could be approximated by,
\begin{equation}\label{eq15}
\begin{array}{lcl}
\delta\varepsilon&=&\sqrt{\delta A^2+\delta S_0^2}\\
&\approx&\delta e\sqrt{\epsilon^2_{A_1}+\epsilon'^2_{A_1}+(\epsilon^2_{E}+\epsilon'^2_{E})\frac{8\Upsilon^2}{4\Upsilon^2+\lambda_{\parallel}^2}},
\end{array}
\end{equation}
where $S_0$ is the zero-field splitting, and the variation in the SO coupling was neglected. Solving for $\delta e$ and substituting in Eq.~(\ref{eq14}) yields,
\begin{equation}\label{eq16}
\delta S\approx\frac{\kappa}{\sqrt{1+\frac{8\kappa^2\Upsilon^2}{4\Upsilon^2+\lambda_{\parallel}^2}}}\delta\varepsilon,
\end{equation}
where $\kappa\equiv\sqrt{(\epsilon^2_E+\epsilon'^2_E)/(\epsilon^2_{A_1}+\epsilon'^2_{A_1})}=1.34$. In the case where the $4\Upsilon^2\gg\lambda^2$, this expression can be further simplified,
\begin{equation}\label{eq16a}
\delta S\approx\frac{\kappa}{\sqrt{1+2\kappa^2}}\delta\varepsilon.
\end{equation}
We have measured the low temperature (5.7~K) inhomogeneous broadening in an as grown optical grade CVD diamond sample (Element Six) to be \mbox{$\delta\varepsilon\approx16$~GHz} (see Appendix~\ref{app:1}). According to \mbox{Eq.~(\ref{eq16a})}, this sets \mbox{$\delta S\approx10$~GHz}.

As this is significantly lower than a reasonably achievable field-induced ground-state splitting ($S\approx50$~GHz, see Sect.~\ref{sect:3} above), spin-echo pulse sequences could indeed be used to counteract the dephasing caused by this broadening. These pulses could be either picosecond millimeter-wave pulses,~\cite{ESRmm1,ESRmm2} or picosecond optical pulses utilizing two-photon transitions.~\cite{Yamamoto08,Kodriano12}

\subsubsection{Effect on polarization selection rules}
Random local strain may also affect the polarization selection rules through its effect on the ground-state wavefunctions. This may weaken the noise suppression effect induced by the cross-linear polarization selection rules achieved with the application of external electric field [Eq.~(\ref{eq9}), Fig.~\ref{fig2}(d)]. Using the eigenstates of the total Hamiltonian, including strain, one can re-derive the noise suppression factor. In the limit of an electric-field induced splitting much larger than S$_0$ and any random-strain energy, this probability now reads,
\begin{equation}\label{eq17}
P\approx\frac{C^2+D^2}{4B^2}.
\end{equation}
For $\Upsilon_y\neq0$, the width of the distribution of this value induced by random strain is (to first order),
\begin{equation}\label{eq18}
\delta P\approx\frac{C\delta C}{2B^2}=\frac{\Upsilon_y\delta S}{S^2},
\end{equation}
where the variation of the SO energy was neglected. Substituting \mbox{$S=50$~GHz} and \mbox{$\delta S=10$~GHz}, one obtains that the maximum upper standard deviation value of the noise suppression factor, $P+\delta P$, is still about 1/10. That is, even in the presence of  random strain on top of the DJT interaction, the suppression of noise through cross-polarization selection rules remains effective.

\subsection{Dynamic distortions: memory lifetime}\label{sect:4b}
Dynamical strain, \emph{i.e.} vibrations, can also couple to the energy levels via the strain Hamiltonian [Eq.~(\ref{eq11})]. This will induce transitions between the ground-states, limiting their lifetime. Since this limitation cannot be circumvented using pulse sequences, it is most important to estimate the expected lifetime due to dynamical distortions. In order to do so, one quantizes the strain in the Hamiltonian of Eq.~(\ref{eq11}), and uses standard time-dependent perturbation theory to derive the transition rate.~\cite{Jelezko_SiV_lifetime} As two-phonon processes are quenched with respect to the single-phonon process at a temperature of a few Kelvin,~\cite{Jelezko_SiV_lifetime} we focus on the single-phonon process. For this process, the rate is given by,~\cite{Jelezko_SiV_lifetime}
\begin{equation}\label{eq19}
\gamma=\frac{2}{\tau}\approx\frac{2\pi}{\hbar}\rho\chi S^3[2N(T,S)+1],
\end{equation}
where $\rho$ is related to the density of states of acoustic phonons in diamond (inversely proportional to the cube of the speed of sound in diamond), $\chi$ is proportional to $\epsilon^2_E+\epsilon'^2_E$, $S$ is the ground-state splitting, and \mbox{$N(T,S)=[\exp(S/k_BT)-1]^{-1}$}, where $k_B$ is the Boltzmann constant, is the average acoustic phonon number at energy $S$ and temperature $T$. Note that both upwards and downwards transitions were included. In Ref.~\onlinecite{Lukin14}, the coherence time of the excited E$_x$ and E$_y$ states of the NV$^-$ is measured at various temperatures for an E$_x$--E$_y$ splitting of \mbox{$S_{NV^-}=3.9$~GHz}. From a global fit to the data, the authors concluded that at the lowest temperature they could measure at, \mbox{$T_{NV^-}=5.8$~K}, the decay rate is consistent with 0 (-0.34$\pm$1.87~MHz, 95\% confidence interval). In order to obtain a lower bound on the life time of the NV$^0$ ground states, we take the upper value of the rate measured in Ref.~\onlinecite{Lukin14}, 1.53~MHz, which yields a minimal life time of \mbox{$\tau_{NV^-}^{min}\approx1.3~\mu$s.} Given this value, the expected lifetime of the NV$^0$ ground states, $\tau_{NV^0}$, can be bounded by,
\begin{equation}\label{eq20}
\tau_{NV^0}>\tau_{NV^-}^{min}\frac{\chi_{NV^-}S_{NV^-}^3[2N(T_{NV^-},S_{NV^-})+1]}{\chi_{NV^0} S_{NV^0}^3[2N(T_{NV^0},S_{NV^0})+1]}.
\end{equation}
The ratio $\chi_{NV^-}/\chi_{NV^0}$ can be estimated using the strain energies of the NV$^0$ and NV$^-$ to be $\sim$0.5.~\cite{DaviesHamer76,Davies79} With this ratio, for $S_{NV^0}$ of 50~GHz, one obtains \mbox{$\tau_{NV^0}>6$~ns} for 4.2~K, and \mbox{$\tau_{NV^0}>17$~ns} for 1~K.
Indeed, optical transition line-widths as low as a few hundred~MHz, comparable to the radiative line-width \mbox{($\sim$100~MHz~\cite{NV0_lifetime})}, have been recently measured for the ZPL of a single NV$^0$ at 2~K.~\cite{Wrachtrup13}
This estimated minimal lifetime is also quite similar to the ground-state lifetime of \mbox{40~ns} recently measured at 4.5~K and zero external fields for the negatively charged silicon-vacancy center, a diamond defect with E symmetry ground states and zero-field splitting of $\sim50$~GHz.~\cite{Jelezko_SiV_lifetime}

\subsection{Spin fluctuations}
Since the initial spin-state is spatially random, it may be possible that due to an effective transverse SO interaction [see Eq.~(\ref{eq5})], possibly induced by mixing with other orbitals,~\cite{Hollenberg11,Maze11} the energy difference between the two memory states will spatially fluctuate. However, as this transverse SO interaction must be less than the transverse SO matrix element, estimated to be on the order of a few GHz,~\cite{Hollenberg11,Maze11} these spatial fluctuations will be a small addition to those induced by random local strain (Sect.~\ref{sect:4a}), and could therefore be dealt with in the same way, using spin-echo techniques.
Such spin-induced random energy changes may also fluctuate with time, due to fluctuating magnetic fields. However, as the time-scale of such fluctuations, due mostly to nuclear spin fluctuations, is on the order of at least microseconds,~\cite{Wrachtrup09} slower than the phonon-induced decay rate discussed in Sect.~\ref{sect:4b} above, they would still be amenable to spin-echo techniques.

\subsection{Intersystem crossing and charge fluctuations: the effect of ``dark states''}
Upon resonant excitation, either to the ZPL, or to the phonon side band, the NV$^0$ may change its charge state into NV$^-$.~\cite{Acosta12} It also has some branching ratio for intersystem crossing to the $^4$A$_2$ low-lying excited states.~\cite{NV0_EPR,Theor_NV0_4A2} Both of these two effects transform the state of the system into a ``dark state''. That is, they quench the optical activity of the NV center at and near the NV$^0$ ZPL, and may therefore limit the memory time. A fluctuating local charge environment may also cause the ground-state energies of the NV$^0$s to fluctuate, and contribute to decoherence. However, as the control and signal fields are nonresonant, they will not transfer population to the NV$^0$ excited states, and will not be able to induce either charge fluctuation or intersystem crossing. These effects may be induced only during optical pumping. However, since the time-scales related to these effects~\cite{NV0_EPR,Acosta12} are much longer than the time scale of optical pumping (the radiative lifetime), they should not considerably interfere with that process as well. Furthermore, optical charging may be counteracted by resonant excitation at the NV$^-$ ZPL,~\cite{Acosta12} thereby improving the efficiency of the optical pumping.

\section{Conclusion}\label{sect:6}
The use of the orbital ground states of the neutral NV center in diamond as the storage states in a quantum optical memory has been proposed and discussed. It was shown that an ensemble such centers could serve as an efficient, low-noise, ultra-broadband quantum-optical memory. The main feature that enables these capabilities is the direct coupling of both ground states to a third, excited level, in orthogonal polarizations. The factors that affect the lifetime of the ground states, namely, random local strain, lattice vibrations, spin fluctuations, and charge fluctuations, have been analyzed. A lifetime of at least a few nanoseconds, long enough to allow for electronic coordination and feed-forward, therefore enabling scalability, is estimated for existing diamond samples at liquid helium temperatures, with the application of spin-echo pulse-trains. The implementation of this proposal may therefore open the way to ultrafast quantum opto-electronic networks in the solid state.

\begin{acknowledgements}
This work was supported by the UK EPSRC (EP/J000051/1), the EU IP SIQS (600645), the US AFOSR EOARD (FA8655-09-1-3020), a Royal Society University Research Fellowship (to JN), and EU Marie-Curie Fellowships (PIEF-GA-2013-627372 to EP, and PIIF-GA-2013-629229 to DJS). IAW acknowledges an ERC Advanced Grant and an EPSRC Programme Grant. EP would like to thank the British Technion Society for a Coleman-Cohen Fellowship, and the Oxford Martin School for initial support.
\end{acknowledgements}

\appendix

\section{Raman coupling mediated by multiple excited states}\label{sect:2}
\subsection{General considerations}
Here we formulate a condition for the first-order Raman coupling (FORC) to vanish identically by destructive interference of contributions from multiple excited states.

Defined using second-order perturbation theory valid for a detuning much larger than the spontaneous decay rate, the Raman coupling coefficient between two different ground-states $i$ and $f$ reads,
\begin{equation}\label{eq1}
R_{fi}^{\beta\alpha}=\sum_k{\frac{D_{fk}^{\beta}D_{ik}^{\alpha*}}{\Delta_k}}\approx\frac{\sum_k{D_{fk}^{\beta}D_{ik}^{\alpha*}}}{\overline{\Delta}}-\frac{\sum_k{D_{fk}^{\beta}D_{ik}^{\alpha*}\varsigma_k}}{\overline{\Delta}^2},
\end{equation}
where $D_{ik}^{\alpha}$ ($D_{fk}^{\beta}$) is the transition dipole moment between excited state $k$ and ground state $i$ ($f$) in polarization $\alpha$ ($\beta$), $\Delta_k$ is the detuning of the carrier frequencies of the two coupling optical fields from single-photon resonances with the $i\leftrightarrow k$ and $f\leftrightarrow k$ transitions, $\overline{\Delta}$ is $\Delta_k$ averaged over all excited states, and \mbox{$\varsigma_k=\Delta_k-\overline{\Delta}$}. The approximation on the right hand side holds for detunings larger than all values of $\varsigma_k$ for which \mbox{$D_{fk}^{\beta}D_{ik}^{\alpha*}\neq 0$}.

The leading term of this approximation, proportional to 1/$\overline{\Delta}$, is the FORC. It is generally the dominant term, and in the case of a single excited state, it is the only term. However, since its numerator is a scalar product of two different rows of the dipole transition matrices $D^{\alpha}$ and $D^{\beta}$, it will completely vanish if at least one of these rows is all zeros, or if the two rows are orthogonal.

If, additionally, the row product in the FORC numerator is zero for \emph{every} two different rows (ground states) for \emph{all} polarizations, while the same-row products assume the same value for a given pair of polarizations, $\{\beta,\alpha\}$, that is, if
\begin{equation}\label{eq2}
D^{\beta}D^{\alpha\dag}=a_{\beta\alpha}I,
\end{equation}
where $I$ is a unit matrix and $a_{\beta\alpha}$ are constants, the FORC will vanish \emph{identically}.

That is, in that case the following two statements apply: (a) From linearity,  \mbox{$D^{\beta'}D^{\alpha'\dag}=a'_{\beta'\alpha'}I$}, for any polarizations $\alpha'$ and $\beta'$. (b) From unitarity, for any \mbox{$\widetilde{D}^{\alpha(\beta)}=UD^{\alpha(\beta)}V^{\dag}$}, where $U$ ($V$) is a unitary matrix operating in the ground (excited) state manifold, \mbox{$\widetilde{D}^{\beta}\widetilde{D}^{\alpha\dag}=a_{\beta\alpha}I$}.

This means that if Eq.~(\ref{eq2}) holds in some state and polarization bases, and therefore the FORC vanishes in that case, the FORC will vanish for \emph{any other} state and polarization bases, as long as ground and excited states are not mixed. Since fields weaker than the ground-excited energy difference transform the ground and excited state manifolds under separate unitary transformations, such fields will not be able to restore FORC in cases where Eq.~(\ref{eq2}) holds.

In cases where the FORC vanishes, the Raman coupling would decay as the average detuning \emph{squared}, and would rapidly become ineffective for detunings larger than the relevant excited state splittings.

\subsection{Coupling between states of different spins}
One important set of cases in which Eq.~(\ref{eq2}) holds and the FORC vanishes identically, involves ground states with different spins (in the term ``spin'' we refer here to both electronic and nuclear spin) and the same orbital function, and excited states that can be unitarily transformed into states that are products of an orbital function and a spin function (SO-products). This is because the transition dipole moments in the SO-product basis are composed of a product of some orbital function and a spin delta-function,
\begin{equation}\label{eq3a}
D^{\alpha(\beta)}_{i(f)k}=d^{\alpha(\beta)}_{o_go_k}\delta_{s_{i(f)}s_k},
\end{equation}
where $o_g$ ($o_k$) denotes the orbital of the ground states ($k^{th}$ excited state), $s_{i(f)}$ denotes the spin of the initial (final), ground state, and $s_k$ denotes the spin of the $k^{th}$ excited state. The product of two rows of the dipole matrices would then be given by,
\begin{equation}\label{eq3b}
\sum_k{D_{fk}^{\beta}D_{ik}^{\alpha*}}=\left(\sum_{o_k}{d_{o_go_k}^{\beta}d_{o_go_k}^{\alpha*}}\right)\delta_{s_is_f}=a_{\beta\alpha}\delta_{if}.
\end{equation}
One can clearly see that this satisfies Eq.~(\ref{eq2}).

The excited states that should be included are those that are closest to resonance with the coupling light, and are separated from one another by less than the detuning. The FORC may therefore not vanish if spin mixing interactions (SMIs), which mix SO-products of different spins and energetically separate them, are larger than the detuning.

One prominent example is the D-lines of Alkali atoms, which stem from transitions between a ground $s$ manifold, and two excited $p$ manifolds mixed and separated by the SO interaction. There, FORC between ground states with different spins vanishes for detunings larger than the SO interaction ($\sim$7~THz in Rb; $\sim$17~THz in Cs), but exists for smaller detunings, \emph{e.g.} when each SO-split $p$ manifold, where the states cannot be transformed into SO-products, can be considered separately.

Another example where the FORC vanishes in this way is the charged exciton in a quantum dot, where without external magnetic fields there is no SMI at all, and all the states can be transformed into SO-products. Indeed, nonresonant optical control, based on stimulated Raman transitions, could be demonstrated only when the transverse external magnetic field used for inducing SMI was high enough to force the level splitting to be on the order of the detuning.~\cite{Yamamoto08}

\subsection{NV$^-$ and NV$^0$}
As shown in Fig.~\ref{fig1}(a), the ground-state manifold of the NV$^-$ can be spanned by three SO-products, $\{$A$_{2,0}$,A$_{2,1}$,A$_{2,-1}\}$, all sharing the same orbital part. The excited states can also be spanned by SO-products, $\{$E$_{x,0}$,E$_{x,1}$,E$_{x,-1}$,E$_{y,0}$,E$_{y,1}$,E$_{y,-1}\}$. As shown above, this is sufficient for the FORC between any two ground states to identically vanish for detunings larger than excited state splittings related to spin mixing. The magnitude of such splittings is set in this case mostly by the magnitude of transverse spin-spin interactions, and therefore would be at most $\sim$1.5~GHz.~\cite{Awschalom14}

For the NV$^0$, on the other hand, when substituting the transition dipole matrices, Eq.~\ref{eq8}, into the left hand side of Eq.~\ref{eq2}, one can immediately see that the row-product matrices $D^xD^{y\dag}$ and $D^yD^{x\dag}$ are non-diagonal: there are pairs of different ground states (different orbitals, same spin) for which the row-product is non-zero. For those states, the FORC will not vanish, and, in fact, will be the only term.\\

\section{Line shape measurements}\label{app:1}
Here we present low temperature fluorescence spectroscopy measurements of the NV$^0$ ZPL, and place upper limits on the inhomogeneous broadening and the zero field splitting.

The studied sample was an optical grade, type Ib cvd diamond, purchased from Element Six. The sample did not go through any post-growth treatments. Nevertheless, fluorescence from both NV$^-$ and NV$^0$ could be quite easily detected upon excitation with 532 nm laser light (Coherent Verdi V8), focused on the sample by a $\times20$, NA=0.29 long working-distance objective lens (Sigma Koki PAL-20-L). The sample was cooled to 5.7~K using a helium flow cryostat (Oxford Instruments). The fluorescence was collected through the same lens and was spectrally analyzed by a 0.48~m spectrometer (Digikr\"om DK480) equipped with a 1200 groove/mm diffraction grating, and a cooled ccd camera (Andor iXon). The spectral resolution at the NV$^0$ ZPL was determined to be 24.5~GHz full width at half maximum by measuring the spectrum of the excitation laser (which has a single longitudinal mode), and scaling by the ratio of the ZPL and laser wavelengths.
\begin{figure}[tbh]
\includegraphics[width=0.5\textwidth]{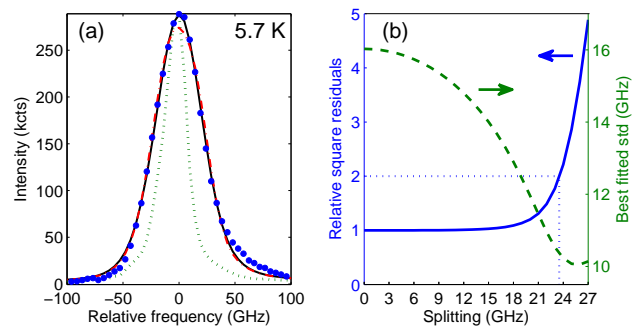}
  \caption{
(a) NV$^0$ ZPL spectrum at 5.7~K. The circles present the measured data. The solid black line is a single Gaussian fit convolved with the measured response function. The latter is presented by the green dotted line. (b) Fitting quality (sum of squared residuals, left axis) and fitted width (one standard deviation, right axis) of a double Gaussian model, versus the splitting between the two Gaussians. The splitting at which the fitting quality is twice worse than the best achievable one is marked by the dotted lines. The corresponding fit is presented in (a) by the red dashed-dotted line.}
\label{fig5}
\end{figure}

The measured spectrum, integrated for 1500~s at 28~mW excitation power, and the measured and scaled response function are presented in Fig.~\ref{fig5}(a) by the blue circles and the green dotted line, respectively. The solid black line is a fit to a single Gaussian convolved with the response function. The standard deviation of the fitted Gaussian is 16~GHz.

In order to estimate an upper limit on a possible zero-field splitting of the ZPL, a two-Gaussians model was fitted to the data. The solid blue line on the left axis of Fig.~\ref{fig5}(b) presents the relative sum of squared residuals of the fit versus the splitting between the two Gaussians. The dashed green line on the right axis presents the best fitted Gaussian width (one standard deviation). An upper limit on the splitting can be placed by the splitting for which the sum of squared residuals doubles with respect to its minimal value (achieved for zero splitting). This occurs for a splitting of $\sim$24~GHz. The corresponding fit is presented in Fig.~\ref{fig5}(a) by the red dashed-dotted line. We therefore set an upper bound of 12~GHz on the DJT energy.


\begin{thebibliography}{43}
\bibitem[{com()}]{comment1}
\bibinfo{note}{Throughout this manuscript we use the convdentional
  crystalographic notation of direct (square brakets) and reciprocal (round
  brakets) lattice vectors. A bar above a number indicates a minus sign.}
\end{thebibliography}
\end{document}